\documentclass[preprint,showpacs,preprintnumbers,amsmath,amssymb]{revtex4}

\usepackage{color}
\usepackage{graphicx}
\usepackage{dcolumn}
\usepackage{bm}
\usepackage{subfigure}

\begin{document}


\title{Revised spherically symmetric solutions of $R+\varepsilon/R$ gravity}

\author{Kh. Saaidi}
 \email{ksaaidi@uok.ac.ir}
\author{S. W. Rabiei}%
 \email{w.rabiei@uok.ac.ir}
 \affiliation{%
Physics Department, University of Kurdistan\\
Sanandaj, Iran
               }%
\author{A. Aghamohammadi}%
 \email{a.aghamohammadi@uok.ac.ir}
\affiliation{%
Islamic Azad University - Sanandaj Branch\\
Sanandaj, Iran
               }%

\date{\today}

\begin{abstract}
We study spherically symmetric static empty space solutions in $R+\varepsilon/R$ model of $f(R)$ gravity. We show that the Schwarzschild metric is an exact solution of the resulted field equations and consequently there are general solutions  which {are perturbed Schwarzschild metric and viable for solar system.
Our results for large scale contains a logarithmic term with a coefficient producing a repulsive gravity force which is in agreement with the positive acceleration of the universe.}
\end{abstract}

\pacs{ }



\maketitle

\section{Introduction}
The current accelerated expansion of the universe is one of the biggest surprises in cosmology \cite{Nr1}. This acceleration may be due to unknown energy momentum components by the equation of the state $p=-\rho$ or some other mechanism. On the other hand it may be justified by modification of general relativity. The idea to modification of general relativity is fruitful and economic with respect to several attempts made to  give explanation for this accelerated expansion of the universe and shortcomings of standard general general relativity \cite{Nr2,Nr3,Nr4,Nr5, rev1}. Relaxing the hypothesis that gravitational lagrangian has to be a linear function of the Ricci curvature scalar, $R$, as in the Hilbert-Einstein formulation, one can take into account, an effective action where the gravitational Lagrangian is a generic $f(R)$ function. One of the initiative $f(R)$ models supposed to explain the positive acceleration of expanding universe has $f(R)$ action as $f(R)=R-\mu^4/R$ \cite{Nr6}. After proposing the $f(R)=R-\mu^4/R$ model it appeared this model suffer several problems. In the metric formalism, initially Dolgov and Kawasaki
discovered the violent instability in the matter sector \cite{r3}. The analysis of this instability generalized to arbitrary $f(R)$
models \cite{r4,r5} and it was shown that an $f(R)$ model is stable if $d^2 f/dR^2 > 0$ and unstable if $d^2f/dR^2 < 0$. Thus we can
deduce $R-\mu^4 / R$ suffers the Dogalov-Kawasaki instability but this instability removes in the $R+\mu^4 /R$ model, where $\mu^4>0$. Furthermore
one can see in $R-\mu^4 / R$ model the cosmology is inconsistent with observation when non-relativistic matter is present. In fact there is no
matter dominant era \cite{r6,r8}. However a recent study shows the standard epoch of matter domination can be obtained in
the $R+\mu^4 /R$ model  \cite{r8}. {The authors of  \cite{r3} had claimed that the $f(R)=R- \frac{\mu ^4}{R}$ does not pass the solar system tests.} Since the Schwarzschild metric is consistent with observations of the solar system, spherically symmetric solutions of a viable theory of  f(R) gravity for the vacuum should have a satisfactory limit of Schwarzschild metric.
As yet, the discussion on the weak field limit of this $f(R)=R+\mu^4/R$ theory of gravity is done and there are a few papers which
claim different results \cite{  far,  chi, chi1, cli, b21, sot, all, noj}.
In the recent works, authors have discussed  $f(R)=R\pm\mu^4/R$ theory of gravity for the case of weak field limit
and obtained perturbation terms in the metric potential components which are independent from the localized mass as source of
inhomogeneous  curvature and initial value conditions  \cite{R20, R21}. This matter independency appears because only a special solution is considered for the trace
equation. In present work we show that the Ricci scalar depends on the initial value conditions and the metric can be reduced to the
space time of Mincowski as the Central body's mass tends to zero.
{By assuming small Ricci scalar}, the present paper analysis \textbf{includes} both weak field limit and strong gravity regime of spherically symmetric
solutions of $f(R)=R+\varepsilon/R$ gravity model at the metric approach for the vacuum. We have shown that the Schwarzschild
metric is an exact solution of the resulted field equations and consequently there are general solutions which {are perturbed
Schwarzschild metric and viable for solar system.}
\section{Theory and argument}

In this section we investigate spherically symmetric static solutions of the vacuum field equations in $R+\varepsilon/R$ model of $f(R)$ gravity. The action for f(R) gravity is
\begin{eqnarray}
S = \int {d^4 x\,\sqrt { - g} \left( {\frac{f(R)}{2} + \kappa {\cal L}_m } \right)},
\label{eq:1}
\end{eqnarray}
where we have set $K=8\pi G = 1$ and  ${\cal L}_m$ is zero for the vacuum. Master Equations are obtained as
\begin{eqnarray}
G_\mu ^\nu  (1 - \frac{\varepsilon }{{R^2 }}) =  - \frac{1}{3}\delta _\mu ^\nu  R - \nabla _\mu  \nabla ^\nu  (\frac{\varepsilon }{{R^2 }}),
\label{eq:2}
\end{eqnarray}
whit

\begin{eqnarray}
G_\mu ^\nu   = R_\mu ^\nu   - \delta _\mu ^\nu  R/2 .
\label{eq:3}
\end{eqnarray}

Contracting the field equation (\ref{eq:2}) we obtain

\begin{eqnarray}
R/3 + \varepsilon /R =  - \Box(\varepsilon /R^2 ),
\label{eq:4}
\end{eqnarray}
and it will be used to simplify equations (\ref{eq:2}).
The analysis of spherically symmetric static solution can be carried out using Schwarzschild coordinate:
\begin{eqnarray}
d\tau ^2  = A(r)dt^2  - B(r)^{ - 1} dr^2 - r^2 d\Omega ^2.
\label{eq:5}
\end{eqnarray}
Substituting the trace equation (\ref{eq:4}) into equations (\ref{eq:2}) give rise to the following equations:

\begin{subequations}
\label{eq:6}
\begin{eqnarray}
R_r^r  &=& \frac{{\varepsilon R + R^3 }}{{(2R^2  - 2\varepsilon )}}
\label{subeq:6a}
\\
R_t^t  &=& R_\theta ^\theta
\label{subeq:6b}
\\
R_\theta ^\theta   &=& R_\varphi ^\varphi
\label{subeq:6c}
\end{eqnarray}
\end{subequations}
where the last equation is trivial.
\par
At first, it is obvious that taking $R=0$ leads to $R_\mu ^\nu   = 0$. Since, equations (\ref{eq:6}) have exact solutions of the Schwarzschild for the metric components, i.e.
\begin{eqnarray}
A(r)=B(r)=1-\frac{2M}{r}
\label{eq:7}
\end{eqnarray}
where $M$ is conventionally a constant positive real number.

Now, we seek perturbed Schwarzschild solutions for equations (\ref{eq:6}). Hence we perturb $R$ slightly from zero to the extent that
\begin{eqnarray}
\frac{{R^2 }}{{|\varepsilon |}} \ll 1 .
\label{eq:8}
\end{eqnarray}
This can be done by adding perturbations to the initial value conditions of the Schwarzschild in some place, namely in $r = r_0$. In fact, since the first derivative of B and the second derivative of A with respect to $r$ appear in equations \ref{eq:6} we should have three arbitrary constant real numbers in a general solution.  Consequently we can expect that equations (\ref{eq:6}) have a solution which is a perturbation of the Schwarzschild metric in a spatial interval containing $r_0$.
Approximating the right side of equation (\ref{subeq:6a}) under condition (\ref{eq:8}) we have:
\begin{subequations}
\label{eq:9}
\begin{eqnarray}
R_r^r  &=&  - R_t^t
\label{subeq:9a}
\\
R_t^t  &=& R_\theta ^\theta
\label{subeq:9b}
\end{eqnarray}
\end{subequations}
Writing the Ricci tensor components in terms of $A(r)$ and $B(r)$, i.e.
\begin{subequations}
\label{eq:10}
\begin{eqnarray}
R_t^t  &=&  - B\left( {\frac{{A'}}{{2A}}} \right)^2  + (\frac{{4B + rB'}}{{2r}})\frac{{A'}}{{2A}} + \frac{{BA''}}{{2A^2 }},
\label{subeq:10a}
\\
R_r^r  &=& \frac{{B'}}{r} + \frac{{B'A'}}{{4A}} - B\left( {\frac{{A'}}{{2A}}} \right)^2  + \frac{{BA''}}{{2A}},
\label{subeq:10b}
\\
R_\theta ^\theta   &=& \frac{{BA'}}{{2rA}} + \frac{{rB' + 2B - 2}}{{2r^2 }},
\label{subeq:10c}
\end{eqnarray}
\end{subequations}
equations 9 can be separated as follows:
\begin{subequations}
\label{eq:11}
\begin{eqnarray}
B' &=& \frac{{1 - B}}{r}
\label{subeq:11a}
\\
A'' &=& \frac{{A'^2 }}{{2A}} - \frac{{A'}}{{2rB}}(1 + B) + \frac{A}{{r^2 B}}(B - 1),
\label{subeq:11b}
\end{eqnarray}
\end{subequations}
where $'$ denotes derivation with respect to $r$.
The solution to equation (\ref{subeq:11a}) is
\begin{eqnarray}
B(r) = 1 - \frac{{2M}}{r}
\label{eq:12}
\end{eqnarray}
where $M$ is a positive constant expected to be close to mass of the central body.
As we mentioned above, answers for equations (\ref{eq:6}) will be perturbations to Schwarzschild metric by choosing initial value conditions slightly deviated from the Schwarzschilds'. If such a solution exists, we can regard $A(r)$ as B(r) + a(r), where a(r) is very small in comparisons to B(r).  By assuming
\begin{eqnarray}
|a(r)| \ll |B(r)|
\label{eq:13}
\end{eqnarray}
equation \ref{subeq:11b} can be linearized as
\begin{eqnarray}
a'' = \frac{{2M(M - r)}}{{r^2 (r - 2M)^2 }}a - \frac{{(r^2  - 5Mr + 6M^2 )}}{{r(r - 2M)^2 }}a',
\label{eq:14}
\end{eqnarray}
with the following solution
\begin{eqnarray}
a(r) = bB(r) + \gamma \alpha (r),
\label{eq:15}
\end{eqnarray}
where $b$ and $\gamma$ are constant real numbers and
\begin{eqnarray}
\alpha (r) = B(r)Ln\left[ {2\sqrt r  + 2\sqrt {r - 2M} } \right] - \sqrt {B(r)}.
\label{eq:16}
\end{eqnarray}
Finally the metric components are presented as follows
\begin{subequations}
\label{eq:17}
\begin{eqnarray}
B(r) &=& 1 - \frac{{2M}}{r}
\label{subeq:17a}
\\
A(r) &=& (1 + b)B(r) + \gamma \alpha (r)
\label{subeq:17b}
\end{eqnarray}
\end{subequations}
Since $bB(r)$ is an exact solution of equation \ref{subeq:11b} and leads to R=0, then only $\gamma \alpha (r)$  should be regarded as a perturbation to the Schwarzschild metric. Using the obtained answers $R$ is written as
\begin{eqnarray}
R(r) = \frac{\gamma }{{2r^{3/2} \left( {r - 2M} \right)^{1/2} }}.
\label{eq:18}
\end{eqnarray}
As one can see when $r \rightarrow 2M$, $B(r)$   tends to zero faster than $\alpha (r)$ and the absolute value of $R$ increases rapidly. In this case as long as
\begin{eqnarray}
\gamma ^2  \ll B(r),
\label{eq:19}
\end{eqnarray}
solution (\ref{subeq:17b}) is a good approximate answer to equation (\ref{subeq:11b}). But for the obtained answers to be valid, we should also check the assumed condition in equation \ref{eq:8}. By inserting the statement for the curvature scalar (i.e. Eq.\ref{eq:18}) into equation \ref{eq:8} it necessitate
\begin{eqnarray}
\frac{{\gamma ^2 }}{{|\varepsilon |}} \ll 4r^3 \left( {r - 2M} \right).
\label{eq:20}
\end{eqnarray}
If we set $\gamma \neq 0$, equations \ref{eq:19} and \ref{eq:20} predict that the metric field can not be a perturbation to the Schwarzschild on the horizon i.e. $r=2M$.

\par
On the other hand, when $r \rightarrow \infty$  , $\alpha (r)$  tends to infinity  logarithmically and it can't fulfill the assumption of equation(\ref{eq:13}) up to spatial infinity. In fact for $r \gg 2M$,
\begin{eqnarray}
A(r) \approx C_1  + C_2 Ln(r),
\label{eq:21}
\end{eqnarray}
is a pretty good approximate answer to equation \ref{subeq:11b} with $C_1$ and $C_2$ being constant real numbers.
For the latter case, curvature scalar $R$ takes the form
\begin{eqnarray}
R(r) \approx \frac{{C_2 }}{{r^2 A(r)}}\,
\label{eq:22}
\end{eqnarray}
Since for $r \gg 2M$  the curvature scalar tends to zero rapidly, the obtained solutions do not violate the condition $R^2/ | \varepsilon | \ll 1$  and equations (\ref{eq:11}) are valid.

\section{Conclusion}

We studied spherically symmetric static empty space solutions in $R+\varepsilon/R$ model of $f(R)$ gravity. {We have shown that:
\begin{itemize}
\item
The Ricci scalar depends on the initial value conditions  and by setting them to zero the central body's mass vanishes and the metric reduces to the space time of Mincowski.
\item
In the limit $R=0$, solution of the model is Schwarzschild metric
\item
In the $\frac{R^2}{|\epsilon|} \ll 1 $ for the intermediate region, the solution of the model is perturbation to Schwarzschild metric and shows where $r \rightarrow 2M $, the solution of model can not be a perturbation to the Schwarzschild metric. Because the requirement $\frac{R^2}{|\epsilon|} \ll 1 $   is not satisfied.
\item
In the large scale limit we have $B(r) \rightarrow 1$ and $A(r) $ take a logarithmic form with an arbitrary coefficient constant. Whereas the scale of the universe is not infinite, so the logarithm term is definite. As well as in the cosmological scale, gravity force depends on the $C_2$'s sign in which $C_2<0$ generates a repulsive  gravity  force which is in agreement with the positive acceleration of the universe.
\end{itemize}
}


\end{document}